\documentclass[aps,prl,twocolumn,showpacs,preprintnumbers]{revtex4}%
\usepackage{amsmath}
\usepackage{graphicx}
\usepackage{dcolumn}
\usepackage{bm}
\usepackage{subfigure}
\usepackage{amsfonts}
\usepackage{amssymb}%

\newcommand{\bmulticol}{\begin{multicols}{2}\narrowtext}
\newcommand{\emulticol}{\end{multicols}\widetext}

\begin{document}
\title{H-theorem for classical matter around a black hole}
\author{Piero Nicolini\thanks{%
Electronic-mail: Piero.Nicolini@cmfd.univ.trieste.it}$^{1,2,3}$
and Massimo Tessarotto%
\thanks{%
Electronic-mail: Massimo.Tessarotto@cmfd.univ.trieste.it}$^{1,4}$}
\address{$^{1}$Dipartimento di Matematica e Informatica,
Universit\`{a} di Trieste, Italy\\
$^{2}$Dipartimento di Matematica, Politecnico di Torino, Turin,
Italy\\  $^{3}$Istituto Nazionale di Fisica Nucleare, Sezione di
Trieste, Italy\\
$^{4}$Consorzio di Magnetofluidodinamica, Trieste,Italy}
\date{\today }

\begin{abstract}
We propose a classical solution for the kinetic description of
matter falling into a black hole, which permits to evaluate both
the kinetic entropy and the entropy production rate of classical
infalling matter at the event horizon. The formulation is based on
a relativistic kinetic description for classical particles in the
presence of an event horizon. An H-theorem is established which
holds for arbitrary models of black holes and is
valid  also in the presence of contracting event horizons.
\end{abstract}

\maketitle


\noindent The remarkable mathematical analogy between the laws of
thermodynamics and black hole (BH) physics following from
classical general relativity still escapes a complete and
satisfactory interpretation. In particular it is not yet clear
whether this analogy is merely formal or leads to an actual
identification of physical quantities belonging to apparently
unrelated framework. The analogous quantities are
$E\leftrightarrow M$, $T\leftrightarrow\alpha\kappa$ and
$S\leftrightarrow(1/8\pi\alpha)A$, where $A$ and $\kappa$ are the
area and the surface gravity of the BH, while $\alpha$ is a
constant. A immediate hint to believe in the thermodynamical
nature of BH comes from the first analogy which actually regards a
unique physical quantity: the total energy. However, at the
classical level there are obstacles to interpret the surface
gravity as the BH temperature since a perfectly absorbing medium,
discrete or continuum, which is by definition unable to emit
anything, cannot have a temperature different from absolute zero.
An reconciliation was partially achieved by in 1975 by Hawking
\cite{Hawking 1975}, who showed, in terms of quantum particle
pairs nucleation, the existence of a thermal flux of radiation
emitted from the BH with a black body spectrum at temperature
$T=\hbar\kappa/2\pi k_{B} $ ({\it Hawking BH radiation model}).
\noindent The last analogy results the most intriguing, since the
area $A$ should essentially be the logarithm of the number of
microscopic states compatible with the observed macroscopic state
of the BH, if we identify it with the Boltzmann definition. In
such a context, a further complication arise when one strictly
refers to the internal microstates of the BH, since for the
infinite red shift they are inaccessible to an external observer.
An additional difficulty with the identification
$S\leftrightarrow(1/8\pi\alpha)A$, however, follows from the BH
radiation model, since it predicts the existence of contracting BH
for which the radius of the BH may actually decrease. To resolve
this difficulty a modified constitutive equation for the entropy
was postulated \cite{Bekenstein 1973,Bekenstein 1974}, in order to
include the contribution of the matter in the BH exterior, by
setting
\begin{equation}
S^\prime = S+\frac{1}{4}k\frac{c^3 A}{G \hbar }, \label{Beck}
\end{equation}
($S^\prime$ denoting the so-called {\it Bekenstein entropy}) where
$S$ is the entropy carried by the matter outside the BH and
$S_{bh}\equiv\frac{1}{4}k\frac{c^3 A}{G \hbar }$ identifies the
contribution of the BH. As a consequence a generalized second law
\begin{equation}
\delta S^\prime\geq 0 \label{second-law}
\end{equation}
was proposed \cite{Bekenstein 1973,Bekenstein 1974} which can be
viewed as nothing more than the ordinary second law of
thermodynamics applied to a system containing a BH. From this
point of view one notices that, by assumption and in contrast to
the first term $S$, $S_{bh}$ cannot be interpreted, in a proper
sense, as a physical entropy of the BH, since, as indicated above,
it may decrease in time. However, the precise definition and
underlying statistical basis both for $S$ and $S_{bh}$ remain
obscure. Thus a fundamental problem still appears their precise
estimates based on suitable microscopic models. Since the
evaluation of $S_{bh}$ requires the knowledge of the internal
structure of the event horizon (excluding for causality the BH
interior), the issue can be resolved only in the context of a
consistent formulation of quantum theory of gravitation
\cite{Bekenstein 1975,Hawking 1976}. This can be based, for
example, on string theory \cite{Reviews 1998} and can be
conveniently tested in the framework of semiclassical gravity
\cite{Nicolini 2001,Nicolini 2002}. Regarding, instead the entropy
produced by external matter $S$, its evaluation depends on the
nature, not yet known, of the BH. However, even if one regards the
BH as a purely classical object surrounded by a suitably large
number of classical particles its estimate should be achievable in
the context of classical statistical mechanics.

In statistical mechanics the ``disorder'' characterizing a
physical system, classical or quantal, endowed by a large number
of permissible microstates, is sometimes conventionally measured
in terms of the so-called Boltzmann entropy $S_B=K\ln W.$ Here $K$
is the Boltzmann constant while $W$ is a suitable real number to
be identified with the total number of microscopic complexions
compatible with the macroscopic state of the system, a number
which generally depends on the specific micromodel of the system.
Therefore, paradoxically, the concept of Boltzmann entropy does
not rely on a true statistical description of physical systems,
but only on the classification of the internal microstates
(quantal or classical).  As is well known\cite{Chakrabarti 1976},
$S_B$ can be axiomatically defined, demanding (i) that it results
a monotonic increasing function of $W$ and (ii) that it satisfies
the entropy additivity law $S_B(W_1 W_2)=S_B(W_1)+S_B(W_2)$.
Boltzmann entropy plays a crucial role in thermodynamics where (i)
and (ii) have their corresponding laws in the entropy
nondecreasing monotonicity and additivity. Since in statistical
mechanics of finite system it is impossible to satisfy both laws
exactly, the definition of $S_B$ is actually conditioned by the
requirement of considering systems with $W>>1$ (large physical
systems).

An alternate definition of entropy in statistical mechanics is the
one given by the Gibbs entropy, in turn related to the concept of
Shannon information entropy. In contrast to the Boltzmann entropy,
this is based on a statistical description of physical systems and
is defined in terms of the probability distribution of the
observable microstates of the system. In many cases it is
sufficient for this purpose to formulate a kinetic description,
and a corresponding kinetic entropy, both based on the
one-particle kinetic distribution function. In particular, this is
the case of classical many-particle systems, consisting of weakly
interacting ultra relativistic point particles, such as those
which may characterize the distribution of matter in the immediate
vicinity of the BH exterior.

The goal of the paper is to to provide an explicit expression for
the contribution $S$, which characterizes Bekenstein law
(\ref{Beck}), to be evaluated in terms of a suitable kinetic
entropy, and to estimate the corresponding entropy production rate
due to infalling matter at the BH event horizon. In addition we
intend to establish an H-theorem for the kinetic entropy which
holds, in principle, for a classical BH characterized by event
horizons of arbitrary shape and size and even in the presence of
BH implosions or slow contractions.  This is obtained in the
framework where the classical description of outside matter and
space is a good approximation to the underlying physics. The basic
assumption is that the matter falling into the BH is formed by a
system of $N\gg 1$ classical point particles moving in a classical
spacetime. We adopt a covariant kinetic formalism taking into
account the presence of an event horizon and assuming Hamiltonian
dynamics for the point particles. The evolution of such a system
is well known and results uniquely determined by the classical
equations of motion, defined with respect to an arbitrary observer
$O$. To this purpose let us choose $O$, without loss of
generality, in a region where space time is (asymptotically) flat,
endowing  with the proper time $\tau$, with $\tau$ assumed to span
the set $I\subseteq R$ (observer's time axis). Each point particle
is described by the canonical state $\mathbf{x}$ spanning the
$8-$dimensional phase space $\Gamma,$ where ${\mathbf{x}}=\left(
r^{\mu},p_{\mu }\right) $. Moreover, its evolution is prescribed
in terms of a suitable relativistic Hamiltonian $H=H\left(
\mathbf{x}\right) $, so that the canonical state
${\mathbf{x}}=\left( r^{\mu}, p_{\mu}\right) $ results
parameterized in terms of the world line arc length $s$ (see
\cite{Synge 1960}). As a consequence, requiring that $s =s(\tau)$
results a strictly monotonic function it follows that, the
particle state can be also parameterized in terms of the
observer's time $\tau.$ To obtain the a kinetic description for a
relativistic classical system of $N$ point particles we
introduce the kinetic distribution function for the observer $O$, $\rho_{G}%
(\mathbf{x})$, defined as follows
\begin{equation}
\rho_{G}({\mathbf{x}})\equiv\rho({\mathbf{x}}) {\displaystyle}
\delta(s-s(\tau)) {\displaystyle} \delta (\sqrt{u_{\mu}u^{\mu}}-1)
\end{equation}
where $\rho\left({\bf x}\right)$ is the conventional kinetic
distribution function in the $8-$dimensional phase space. Notice
that the Dirac deltas here introduced must be intended as
\emph{physical realizability equations}. In particular the
condition placed on the arc length $s$ implies that the particle
of the system is parameterized with respect to $s({\tau})$, i.e.,
it results functionally dependent on the proper time of the
observer; instead the constraints placed on $4$-velocity implies
that $u^{\mu}$ is a tangent vector to a timelike geodesic. The
\emph{event horizon} of a classical BH is defined by the
hypersurface
 $r_{H}$ specified by the equation
\begin{equation}
R(x)=r_{H}%
\end{equation}
where $x$  denotes a point of the space time manifold, while
$R(x)$ reduces to the radial coordinate in the spherically
symmetric case. According to a classical point of view, let us now
assume that the particles are "captured" by the BH (i.e., for
example, they effectively disappear for the observer since their
signals are red shifted in such a way that they cannot be anymore
detected {\cite{Wald 1984}}) when they reach $\gamma$ of equation
\begin{equation}
R_\epsilon (x)= r_{\epsilon}.
\end{equation}
Here $r_\epsilon=(1+\epsilon )r_H$, while $\epsilon$ depends on
the detector and the $4-$momentum of the particle. The presence of
the BH event horizon is taken into by defining suitable boundary
conditions for the kinetic distribution function on the
hypersurface $\gamma $. For this purpose we distinguish between
incoming and outgoing distributions on $\gamma$,
$\rho_{G}^-(\mathbf{x})$ and $\rho_{G}^+(\mathbf{x})$
corresponding respectively to $n_\alpha u^\alpha \leq 0$ and
$n_\alpha u^\alpha > 0$, where $n_\alpha$ is a locally radial
outward $4-$vector. Therefore the boundary conditions on $\gamma$
are specified as follows
\begin{eqnarray}
&&\rho_{G}^+({\mathbf{x})}\equiv\rho({\mathbf{x}})%
{\displaystyle}
\delta(s-s(\tau))%
{\displaystyle}
\delta(\sqrt{u_{\mu}u^{\mu}}-1)\\
&&\rho_{G}^-({\mathbf{x}})\equiv0%
\end{eqnarray}
It follows that it is possible to represent the kinetic
distribution function in the whole space time manifold in the form
\begin{equation}
\rho_G\left(\mathbf{x}\right)=\rho_G^-\left(\mathbf{x}\right)+
\rho_G^+\left(\mathbf{x}\right)
\end{equation}
where
\begin{eqnarray}
\rho_{G}^\pm({\mathbf{x}})&\equiv&\rho({\mathbf{x}})%
{\displaystyle}
\delta(s-s(\tau))%
{\displaystyle} \delta(\sqrt{u_{\mu}u^{\mu}}-1)\times%
{\displaystyle}\nonumber\\
&&\times\Theta^\pm(R_{\epsilon}(x)-r_{\epsilon}(s(\tau)))
\end{eqnarray}
with $\Theta^\pm$ respectively denoting  the strong and the weak
Heaviside functions
\begin{equation}
\Theta^+(a)=\left\{
\begin{array}
[c]{ccc}%
1 & for & a\geq0\\
0 & for & a<0.
\end{array}
\right.
\end{equation}
and
\begin{equation}
\Theta^-(a)=\left\{
\begin{array}
[c]{ccc}%
1 & for & a>0\\
0 & for & a\leq 0.
\end{array}
\right.
\end{equation}
We stress that in the above boundary conditions no detailed
physical model is actually introduced for the particle loss
mechanism, since all particles are assumed to be captured on the
same hypersurface $\gamma$, independent of their mass, charge and
state. This provides a classical loss model for the BH event
horizon.

 Let us now consider the evolution of the the kinetic distribution
function $\rho_{G}(\mathbf{x})$ in external domains, i.e. outside
the event horizon. Assuming that binary collisions are negligible,
or can be described by means of a mean field, and provided that
the phase space volume element is conserved, it follows the
collisionless Boltzmann equation, or the Vlasov equation in the
case of charged particles \cite{Tessarotto2004},
\begin{equation}
\frac{ds}{d\tau}\left\{ \frac{dr^{\mu}}{ds}\frac{\partial
\hat{\rho}(\mathbf{x})}{\partial r^{\mu}}+\frac{dp_{\mu}}{ds%
}\frac{\partial\hat{\rho}(\mathbf{x})}{\partial p_{\mu}}\right\}
=0 \label{Liouville equation}%
\end{equation}
with summation understood over repeated indexes, while
$\widehat{\rho}(\mathbf{x})$ denotes $\rho_G(\mathbf{x})$
evaluated at $r^{0}=r^{0}(s(\tau))$ and $p_{0}=m\left\vert
\frac{\partial r_{0 }(s)}{\partial s}\right\vert _{s=s(\tau)}.$
This equation resumes the conservation of the probability in the
relativistic phase space in the domain external to the event
horizon. Invoking the Hamiltonian dynamics for the system of point
particles, the kinetic equation takes the conservative form
\begin{equation}
\frac{ds}{d\tau}\left[  \hat{\rho}({\mathbf{x}}),H\right]  _{\mathbf{x}%
}=0 .
\end{equation}
Let us now introduce the appropriate definition of kinetic entropy
$S(\rho)$ in the context of relativistic kinetic theory. We intend
to prove that in the presence of the BH event horizon it satisfies
an $H$ theorem. The concept of entropy in relativistic kinetic
theory can be formulated by direct extension of customary
definition given in nonrelativistic setting \cite{Grad 1956,Israel
1963,Cercignani 1975,De Groot 1980}. For this purpose we introduce
the notion of kinetic entropy measured with respect to an observer
endowed with proper time $\tau$ as follows
\begin{equation}
S(\rho)=-%
P{\displaystyle\int\limits_{\Gamma}}
d{\mathbf{x}}(s)
{\displaystyle}
\delta(s-s(\tau))%
{\displaystyle}
\delta(\sqrt{u_{\mu}u^{\mu}}-1)\rho({\mathbf{x}})\ln\rho({\mathbf{x}})
\label{Kentropy}
\end{equation}
where $\rho(\mathbf{x})$ is strictly positive and, in the
8-dimensional integral, the state vector ${\bf x}$ is
parameterized with respect to $s$, with $s$ denoting an arbitrary
arc length. Here $P$ is the principal value of the integral
introduce in order to exclude from the integration domain the
subset in which the distribution function vanishes. Hence
$S(\rho)$ can also be written:
\begin{equation}
S(\rho)=- P
{\displaystyle\int\limits_{\Gamma}}
d{\mathbf{x}}(s)
{\displaystyle}
\delta(s-s(\tau))%
{\displaystyle}
\rho_1({\mathbf{x}})\ln\rho(\mathbf{x}),
\end{equation}
where $\rho_{1}(\mathbf{x})$ now reads%
\begin{equation}
\rho_{1}({\mathbf{x}})=
{\displaystyle}
\Theta(r(s)-r_\epsilon
(s))\delta(\sqrt{u_{\mu}u^{\mu}}-1)\rho({\mathbf{x}}(s)).
\end{equation}
Differentiating with respect to $\tau$ and introducing the
invariant volume element $d^3{\bf r}d^3{\bf p}$, the entropy
production rate results manifestly proportional to the area $A$ of
the event horizon
\begin{equation}
\frac{dS(\rho)}{d\tau}=-%
P{\displaystyle\int\limits_{\Gamma^-}}
d^3{\mathbf{r}}d^3{\mathbf{p}}
{\displaystyle}
F_{{ r}{r_\epsilon}}
\left[\delta\left(r-r_\epsilon\right)\hat{\rho} \ln\hat{\rho}
\right]. \label{entro}
\end{equation}
Indeed, the r.h.s represents the entropy flux across the event
horizon.  Moreover here, $\Gamma^-$ is the subdomain of phase
space corresponding to the particle falling into the BH and $F_{{
r}{r_\epsilon}}$ is the characteristic integrating factor
\begin{equation}F_{{ r}{r_\epsilon}}\equiv\frac{ds(\tau)}{d\tau}%
\left(\frac{dr}{ds}-\frac{dr_\epsilon}{ds}\right).\end{equation}
We can write the above expression in terms of the kinetic
probability density evaluated at the hypersurface $\sqrt{u_\mu
u^\mu -1}$, defined as $\hat{f}({\bf x})\equiv \hat{\rho}/N$. It
follows
\begin{equation}
\hat{\rho}\ln\hat{\rho}\equiv N\hat{f}\ln N\hat{f}.
\end{equation}
At this point we adopt a customary procedure in statistical
mechanics \cite{Yvon 1969} invoking the inequality
\begin{equation}
N\hat{f}\ln  N\hat{f}\geq N\hat{f}-1
\end{equation}
and notice that in the subdomain of phase space $\Gamma^-$ in
which $F_{{ r}{r_\epsilon}}\ge 0$ there results by definition
$\hat{\rho}=0$. Hence it follows that in $\Gamma^-$
\begin{equation}
F_{{ r}{r_\epsilon}}< 0
\end{equation}
where by construction  $\frac{ds(\tau)}{d\tau}> 0$. This result
holds independent of the value of  $\frac{dr_\epsilon}{ds}$. Next
let us introduce  the bounded subset $\Omega \subset \Gamma^-$
such that $N\hat{f}$ results infinitesimal (of order $\delta$ ) in
the complementary set $\Gamma^- \setminus\Omega$ and moreover the
ordering estimate
\[ P {\displaystyle\int\limits_{\Gamma^-
\setminus\Omega}} d^3{\mathbf{r}}d^3{\mathbf{p}}
{\displaystyle}\left|F_{{ r}{r_\epsilon}}\right|
\delta\left(r-r_\epsilon\right) \left[N\hat{f}\ln
N\hat{f}\right]\sim O\left(\delta \ln\delta\right)  \] is required
to hold. Manifestly this domain includes the set of improper
points of $\Gamma^-$. In the set $\Omega$, $N\hat{f}$ is by
assumption positive and such that $N\hat{f}>\delta$. Therefore
there results
\begin{equation}
0< P {\displaystyle\int\limits_{\Omega}}
d^3{\mathbf{r}}d^3{\mathbf{p}} {\displaystyle}\left|F_{{
r}{r_\epsilon}}\right| \delta\left(r-r_\epsilon\right)\equiv
M_\delta
\end{equation}
where $M_\delta$ is a suitable finite constant. Thus one obtains
\begin{equation}
\frac{dS(\rho)}{d\tau}\geq P {\displaystyle\int\limits_{\Omega}}
d^3{\mathbf{r}}d^3{\mathbf{p}} {\displaystyle}\left|F_{{
r}{r_\epsilon}}\right| \delta\left(r-r_\epsilon\right)\left[ Nf-1
\right]+O\left(\delta \ln\delta\right) \label{vel}
\end{equation}
The first term of the r.h.s  of (\ref{vel}) can be interpreted in
terms of the effective radial velocity of incoming particles
\begin{equation}
V_{r}^{eff}\equiv\frac{1}{n_0}{\displaystyle\int\limits_{\Omega}}
d^3{\mathbf{p}} {\displaystyle}\left|F_{{ r}{r_\epsilon}}\right|
\delta\left(r-r_\epsilon\right)\hat{f},
\end{equation}
while $n_0$ is the surface number density of the incoming particle
distribution function
\begin{equation}%
{\displaystyle\int\limits_{\widehat{\Omega}}}
d^{3}{\mathbf{r}}d^{3}{\mathbf{p}}\hat{f}({\mathbf{x}})\delta(r%
-r_{0}(s(t))= An_{o}.
\end{equation}
Finally we invoke the
majorization  %
\begin{eqnarray}
\frac{dS(\rho)}{d\tau}&\geq& \dot{S}\equiv N \inf
\left\{\int_{\Omega} d^3{\bf r} n_0
V_r^{eff}\right\}\nonumber\\&&-M_\delta +O\left(\delta \ln\delta
\right) \label{essep}
\end{eqnarray}
and impose that $N\gg 1$ be sufficiently large to satisfy the
inequality $\dot{S}> 0.$ We stress that $\inf \left\{\int d^3{\bf
r} n_0 V_r^{eff}\right\}$ can be assumed strictly positive for non
isolated BHs surrounded by matter. This proves the relativistic
H-theorem
\begin{equation}
\frac{dS(\rho)}{d\tau}>0.
\end{equation}
Let us briefly analyze the basic implications of our results.
First we notice that the H-theorem here obtained appears of
general validity even if achieved in the classical framework and
under the customary requirement $N\gg 1$ (large classical system)
and in validity of the subsidiary condition $\dot{S}>0$. Indeed
the result applies to BH having, in principle, arbitrary shape of
the event horizon. The description adopted is purely classical
both for the falling particles (charged or neutral
\cite{Tessarotto1998,Tessarotto1999,Tessarotto2004,Nicolini2005})
and for the gravitational field and is based on the relativistic
collisionless Boltzmann equation and/or the Vlasov equation
respectively for neutral and charged particles. A key aspect of
our formalism is the definition of suitable boundary conditions
for the kinetic distribution function in order to take into
account the presence of the event horizon. Second, the expressions
for the entropy and entropy production rate, respectively
Eqs.(\ref{Kentropy}) and (\ref{entro}), can be used to determine
the Bekenstein entropy for classical BH (\ref{Beck}) and the
related generalized second law (\ref{second-law}), although we
stress that the present results are independent of the assumptions
involved in the definition of the Bekenstein entropy
(\ref{entro}). Finally interesting features of the derivation are
that the entropy production rate results proportional to the area
of the event horizon and that the formalism is independent of the
detailed model adopted for the BH. In particular also the possible
presence of an imploding star (contracting event horizon) is
permitted.

\section*{Acknowledgments}

P. N. is supported by the Ministero dell'Istruzione,
dell'Universit\`{a} e della Ricerca (MIUR) via the Programma PRIN
2004: `Metodi matematici delle teorie cinetiche'.

\end{document}